\newcommand{\be}{\begin{equation}}
\newcommand{\ee}{\end{equation}}
\newcommand{\ber}{\begin{eqnarray}}
\newcommand{\eer}{\end{eqnarray}}

\documentstyle[preprint,epsfig,epsf,aps,floats,psfrag]{revtex}
\begin{document}
\tighten
\preprint{\vbox{
\hbox{INT-PUB 03-08}}}
\bigskip
\title{Charge Neutrality of the Color-Flavor Locked Phase from the Low Energy Effective Theory}
\author {Andrei Kryjevski \footnote{abk4@u.washington.edu}}
\address{Dept. of Physics and Institute for Nuclear Theory, 
University of Washington, Seattle, WA 98195}
\date{\today}
\maketitle

\begin{abstract}

We investigate the issue of charge neutrality of the CFL$K^0$ phase of dense quark matter using the 
low energy effective theory of high density QCD.
We show that the local electric and color charge neutrality of the ground state in a homogeneous color 
superconducting medium follows from its dynamics.
We also consider the situation of a spatially inhomogeneous medium, such as may be found in a
neutron star core. We find that spatial inhomogeneity
results in the generation of electric fields, and positrons and/or electrons may be present in the ground
state. We estimate the concentration of charged leptons 
in the ground state to be $n_{e}\sim{10^2}{cm}^{-3}$ and consider their influence on the opacity of the medium with 
respect to the modified photons.

\end{abstract}
 
   
\section {Introduction}

It is now understood that at asymptotically large baryon number densities, the ground
state of 3-flavor, 3-color massless QCD is the Color-Flavor Locked (CFL) 
phase\cite{arw,schafer_gs}. (See \cite{review1} for comprehensive reviews.) In this phase, all 9 species of quarks with momenta close to 
the Fermi surface undergo BCS-like pairing.  The resulting quark-quark 
condensate spontaneously breaks the original $SU(3)_{C}\times SU(3)_L \times 
SU(3)_R$ symmetry down to the diagonal subgroup, $SU(3)_{C+L+R}$, 
causing the gauge bosons of the original $SU(3)_{C}$ group to 
become massive, and also breaks the $U(1)_B$ symmetry associated with conserved baryon number.\footnote{A local gauge symmetry
cannot really be spontaneously broken \cite{elitzur}. We have to fix a gauge 
to define expectation values of gauge variant quantities.}
One linear combination of gluon and photon remains 
massless; the associated unbroken gauge symmetry is referred to as 
$U(1)_{\widetilde{Q}}$, denoting the ``modified'' electromagnetism.  

A nonzero mass for the strange quark encourages the system
to reduce the strange quark number density relative to the density of up- and
down-type quarks. It has been argued that the system responds to this stress by forming a 
kaon condensate in the ground state \cite{schafer,bed_schafer,kaplanreddy}. Kaon condensation allows the strange quark number density to 
be decreased 
without the costly breaking of pairs in the CFL background.

In particular, it was found that in the presence of small enough
chemical potentials for the electric and lepton charges $K^0$ condensate was present
in the ground state (the CFL$K^0$ phase) \cite{bed_schafer,kaplanreddy}. The CFL and CFL$K^0$ phases are distinct, as hypercharge is 
spontaneously broken 
in the latter, but not in the former. Some recent reviews are listed in \cite{review2}.

Recently, some attention has been devoted to the issue of realization of charge neutrality
in the high baryon density matter. First, it was argued that for a macroscopic system the difference between being in a color singlet 
state and being in 
a state with equal number densities of red, blue and green color charges is negligible \cite{birse}. Alford and Rajagopal constructed 
an expression for the free energy 
that incorporated general features of the CFL phase \cite{alfordrajagopal}, while Steiner, Reddy and Prakash used NJL 
model supplemented by diquark and
the t'Hooft six-fermion interactions \cite{steineretal}. Neither approach included gauge bosons as dynamical fields. 
Instead, to ensure charge neutrality of the CFL phase, chemical potentials $\mu_3$ and $\mu_8$ were introduced for 
the charges corresponding to $T_3$ and $T_8$ generators of the $SU(3)_c$ group, along with $\mu_Q$, a
chemical potential for the electric charge. In both papers it was found that $\mu_3,$ $\mu_8$ and $\mu_Q$ had to be adjusted 
to ensure local charge neutrality of the bulk quark matter. 

Motivated by the ideas of \cite{alfordrajagopal}, in this paper we employ the low energy effective theory of QCD at high baryon number 
density developed in \cite{casalbuoni_gatto,sonsteph,bed_schafer} to 
address the issue of charge neutrality in the high baryon density matter. In particular, in this 
approach gauge fields are treated as dynamical degrees of freedom. One expects that in a color superconducting state, color charge 
neutrality will follow 
from the low energy dynamics of the system, and that is what we find for a homogeneous case. 

In section 2  we show that this is indeed the case for a homogeneous medium; in  section 3  we present the analysis
for the case of an inhomogeneous medium, such as one would expect in a compact star, and we find a surprising result that nonzero electric 
fields
exist in the bulk in this case.

\section{Charge Neutrality in Homogeneous Medium}

QCD at asymptotically high baryon density has several effective field theories valid for different ranges of excitation energies. For 
energies well below 
the quark number chemical potential $\mu$ heavy anti-quarks (mass $\sim 2 \mu$) may be integrated out in a systematic manner leading to the 
High Density Effective Theory (HDET) developed in \cite{hong,beane_savage,bed_schafer}. 
 
For energies smaller than the superconducting gap $\Delta$ quasi-particle and quasi-hole states in the vicinity of the Fermi surface may, 
in turn, be 
integrated out
leaving the Nambu-Goldstone bosons 
due to breaking of various flavor symmetries in the ground state as the relevant degrees of freedom 
\cite{casalbuoni_gatto,sonsteph,bed_schafer,cas_gat_nard}. For instance, in the 
CFL phase we have an octet due to breaking of the chiral symmetry and a singlet due to $U(1)_B$ breaking. See \cite{review3} for a review 
of the subject.The low energy effective theory 
of dense QCD is similar in spirit to the 
effective theory of an ordinary (QED) superconductor (Ginzburg-Landau theory). See, for example, a pedagogical introduction by S. Weinberg 
\cite{weinberg}.

The leading terms of the low energy effective Lagrangian of QCD at high baryon number density are
\ber
{\mathcal L} &=& -{f^2_{\pi}\over{2}} \left[{\mathrm Tr}\left((X^{\dag} D_{0}^{L} X)^2 - {1\over{3}}(X^{\dag} D_{i}^{L} X)^2\right)+
{\mathrm Tr}\left((Y^{\dag} D_{0}^{R} Y)^2 - {1\over{3}}(Y^{\dag} D_{i}^{R} Y)^2 \right)\right] \nonumber \\
&&-{1\over{4}}F_{\mu\nu}F^{\mu\nu}-{1\over{2}}{\mathrm Tr}(G_{\mu\nu}G^{\mu\nu}),
\label{lagrangian}
\eer
where $X$ and $Y$ are the $3 \times 3$ unitary matrix valued composite fields describing oscillations of the left- and right handed 
quark-quark condensates about the CFL ground state. 
Under the original symmetry group $SU(3)_{c}\times SU(3)_{L}\times SU(3)_{R},$ 
X transforms as $(3,3,1)$ and Y as $(3,1,3)$. The pion decay constant $f_{\pi}\simeq 0.209 {\mu}$ 
was computed in \cite{sonsteph}. The gauge covariant 
derivatives are

\ber
&& D_{\nu}^{L} X= \partial_\nu X + i g_s X (A_\nu^{c})^{T} + i e A^{em}_\nu Q X + i \delta_{\nu 0}{{M M^{\dag}}\over{2 \mu}} X,\nonumber \\
   &&
 D_{\nu}^{R} Y= \partial_\nu Y + i g_s Y (A_\nu^{c})^{T} + i e A^{em}_\nu Q Y + i \delta_{\nu 0}{{M^{\dag} M}\over{2 \mu}} Y,
    \label{cov_der}
    \eer
where $A_{\nu}^{c}$ and $A^{em}_{\nu}$ are the gluon and photon gauge fields, respectively; $A_{\nu}^c=A_{a\nu}^c t_a,$ where $t_a$ is an 
$SU(3)_{c}$ generator in the fundamental representation.
We treat quark mass matrix $M$ as an external (spurion) field that 
transforms as $(1,3,\bar{3}).$ It was argued in \cite{bed_schafer} that ${M^2}/{2 \mu}$ terms should appear in the covariant 
derivatives (\ref{cov_der}), and, thus, that in the low energy effective theory ${m_s^2}/{2 \mu}$ plays the role of chemical potential 
for strangeness. These are the terms responsible for the formation of kaon condensate in the ground state.
In this paper we neglect up and down quark masses and set 
\be
M={\mathrm diag}(0,0,m_s).
\label{M}
\ee 
The quark electric charge matrix is $Q={\mathrm diag}(2/3,-1/3,-1/3).$ 

A generic term in the effective Lagrangian is made of various $SU(3)_{C}\times SU(3)_L \times SU(3)_R\times U(1)_B$ invariant combinations 
of powers of
${f^2_{\pi}\Delta^2} Y^{\dag}({D\over{\Delta}})^k Y,$ ${f^2_{\pi}\Delta^2} X^{\dag}({D\over{\Delta}})^n X,$ $G_{\mu\nu}$ and $F_{\mu\nu}.$ 
The terms 
involving both $X$ and $Y$ and the ones containing the field strengths are further suppressed by powers of $g_s$ and/or $e$ \cite{sonsteph}.

It is more convenient to rewrite the Lagrangian in the following form \cite{casalbuoni_gatto}

\ber
{\mathcal L} = {\mathcal L}_{X Y A} + {\mathcal L}_{\Sigma}, 
\label{lagrangian1}
\eer
where
\ber 
{\mathcal L}_{X Y A} &=& -{f^2_{\pi}\over{4}} {\mathrm Tr}\left(X^{\dag} \dot {X} + Y^{\dag} \dot {Y} + 
i X^{\dag} ({{M M^{\dag}}\over{2 \mu}} + e A^{em}_0 Q ) X + i Y^{\dag} ({{M^{\dag} M}\over{2 \mu}} + e A^{em}_0 Q) Y + 
2 i g_s (A_0^{c})^{T}\right)^2 \nonumber \\
&& 
 +{f^2_{\pi}\over{4}}{1\over{3}}{\mathrm Tr} \left( X^{\dag} \nabla_i X +
Y^{\dag} \nabla_i Y + i e A^{em}_i( X^{\dag} Q X + Y^{\dag} Q Y) + 2 i g_s (A_i^{c})^{T}\right )^2  \nonumber \\
&&-{1\over{4}}F_{\mu\nu}F^{\mu\nu}-{1\over{2}}{\mathrm Tr}(G_{\mu\nu}G^{\mu\nu}),
\label{lagrangian_xya}
\eer 
and
\ber
{\mathcal L}_{\Sigma} &=& -{f^2_{\pi}\over{4}}{\mathrm Tr}\left(X^{\dag} \dot {X} - Y^{\dag} \dot {Y}+ i X^{\dag} 
{{M M^{\dag}}\over{2 \mu}} X 
 - i Y^{\dag} {{M^{\dag} M}\over{2 \mu}} Y  + i \tilde{e} A^{\widetilde Q}_0 (X^{\dag} Q X - Y^{\dag} Q Y) \right)^2 \nonumber \\
 && 
+{f^2_{\pi}\over{4}}{1\over{3}}{\mathrm Tr} \left( X^{\dag} \nabla_i X -
Y^{\dag} \nabla_i Y + i \tilde{e} A^{\widetilde Q}_i( X^{\dag} Q X - Y^{\dag} Q Y) \right )^2 \nonumber \\
&=& {f^2_{\pi}\over{4}}{\mathrm Tr}\left( \dot \Sigma + i W_L \Sigma - i\Sigma W_R \right) \left( \dot \Sigma^{\dag} + 
 i W_R \Sigma^{\dag} - i\Sigma^{\dag} W_L \right) \nonumber \\
&& -{f^2_{\pi}\over{4}}{1\over{3}}{\mathrm Tr}\left( \nabla_i \Sigma + i \tilde{e} A^{\widetilde Q}_i [Q,\Sigma]\right) 
\left( \nabla_i \Sigma^{\dag}
 + 
 i \tilde{e} A^{\widetilde Q}_i [Q,\Sigma^{\dag}] \right) .
\label{lagrangian_sigma}
\eer
Here 
\ber 
W_L=\tilde{e} A^{\widetilde Q}_0 Q + {{M M^{\dag}}\over{2 \mu}}
\label{w_l}
\eer
and 
\ber
W_R=\tilde{e} A^{\widetilde Q}_0 Q + {{M^{\dag} M}\over{2 \mu}},
\label{w_r}
\eer
$A^{\widetilde Q}_\mu$ is the $U_{\widetilde Q}(1)$ gauge field with the coupling constant 

\be
\tilde{e}=\frac{\sqrt{3} e g_s}{\sqrt{3 g_s^2 + 4 e^2}}.
\label{tilde_e}
\ee 

We have introduced the color singlet field $\Sigma=X Y^{\dag}={\mathrm exp}[2i\pi^a t^a/f_{\pi}],$
where $\pi_a$'s are the pseudo-scalar octet of Nambu-Goldstone bosons which arise from the breaking of chiral symmetry. $\Sigma$ 
transforms as $(1,3,\bar{3}).$ We neglect the $U_B(1)$ Nambu-Goldstone boson terms in the Lagrangian as they are irrelevant for the gauge 
charge neutrality.

We see that ${\mathcal L}_{\Sigma}$ only depends on light fields, while ${\mathcal L}_{X Y A}$ has both light and heavy degrees of freedom 
in it. The convenience of the representation (\ref{lagrangian1}) is due to the fact that at low energy the heavy gauge fields may be 
integrated out. Then ${\mathcal L}_{X Y A}$
piece vanishes leaving us with  ${\mathcal L}_{\Sigma}$ only.
Now one may proceed to solve equations of motion to determine values of various (classical) fields in the ground state. It was found in  
\cite{bed_schafer,kaplanreddy} that in the ground state $\Sigma$ field takes on a non trivial value 
\be
\Sigma_{K^0} = \left( \begin{array}{ccc}
1 & 0 & 0 \\
0 & 0 & i \\
0 & i & 0 \end{array} \right)
\label{ansatz}
\ee corresponding to CFL$K^0$ phase. We note that in the approximation $m_u=m_d=0$ the $K^0$ 
condensate in the ground state is maximal and 
formed for arbitrarily small value of $m_s$ \cite{kaplanreddy}.

\begin{figure}[t]
\centering{
\epsfig{figure=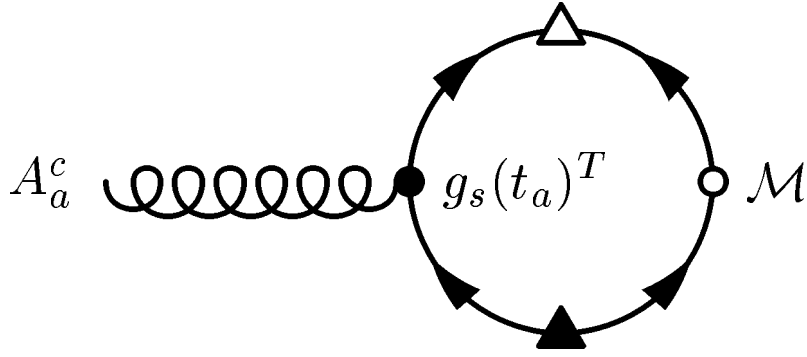, width=.65\textwidth}
}
\caption{This is the HDET diagram that produces operator (\ref{colorcharge}) at one loop level. Fermion lines with triangle insertions 
denote the anomalous
quasi particle propagators. Hollow circle stands for insertion of ${\mathcal M}$
defined in (\ref{mcal_m}). Normal one loop diagram contribution vanishes.}
\label{fig1}
\end{figure}

From ${\mathcal L}_{X Y A}$ we see that 
$X^{\dag} {{M M^{\dag}}\over{2 \mu}} X + Y^{\dag} {{M^{\dag} M}\over{2 \mu}} Y$ acts as a source term for the temporal components of gauge 
fields. This follows from ${M^2}/{2 \mu}$ terms being part of the covariant derivatives (\ref{cov_der}).
We may also see this from the matching calculation of the coefficient of the operator 
\be
g_s A_{a}^{c} {\mathrm Tr} \left( (t_a)^{T}{\mathcal M}\right),
\label{colorcharge}
\ee
where 
\be
{\mathcal M}= X^{\dag} {{M M^{\dag}}\over{2 \mu}} X + Y^{\dag} {{M^{\dag} M}\over{2 \mu}} Y, 
\label{mcal_m}
\ee
in the low energy 
effective Lagrangian (\ref{lagrangian1}). 
The calculation yields a non zero coefficient of $f^2_{\pi}/4$ that comes from the anomalous contribution to the 
one-loop tadpole diagram with one insertion of ${\mathcal M}$ as shown in 
Fig. \ref{fig1}. The diagram is superficially proportional to $\Delta^2,$ but the loop integration is infrared sensitive and picks up a 
compensating $1/{\Delta^2}.$

We note that spatial components of gauge fields do not have sources and vanish in the ground state. 

To proceed 
it is convenient to define a unitary matrix $\xi$, such that $\xi^2=\Sigma$, and an $SU(3)$ matrix $U$. Then $X$ and $Y$ may be presented 
as 
\ber
X&=&\xi U, \nonumber \\
Y&=&\xi^\dag U
\label{newXYparametrization}
\eer

We fix $SU(3)_C$ gauge by setting
\be
U=1,
\label{gaugefix}
\ee
which corresponds to the unitary gauge \cite{casalbuoni_gatto}.
Then we have
\be
\xi_{K^0} ={1\over{\sqrt{2}}} 
\left( \begin{array}{ccc}
\sqrt{2} & 0 & 0 \\
0 & 1 & i \\
0 & i & 1 \end{array} \right)
\label{sigmaK0}
\ee 
and $X=\xi_{K^0}$, $Y=\xi_{K^0}^{\dag}.$

In the homogeneous medium all fields are constants and the Lagrangian reduces to
\ber
{\mathcal L} &=&{f^2_{\pi}\over{4}}{\mathrm Tr}\left({\xi_{K^0}^{\dag}}{{M M^{\dag}}\over{2 \mu}}{\xi_{K^0}} + 
{\xi_{K^0}}{{M^{\dag} M}\over{2 \mu}}{\xi_{K^0}}^{\dag}  
 + 2 g_s (A_0^{c})^{T} +  2 e A^{em}_0 Q \right )^2
\label{lagrangian3}
\eer

Now let us proceed to state the $\it{main}$ result of the paper. The charge densities are defined as 
\be
J_0=-{\partial \over {\partial A_0}}{\mathcal L},
\label{chargedensity}
\ee
while the equations of motion for the temporal components of the gauge fields in the homogeneous case read
\be
{\delta \over {\delta A_0}}{\mathcal L}\equiv{\partial \over {\partial A_0}}{\mathcal L}=0.
\label{eom}
\ee
Then color and electromagnetic charge densities vanish in the ground state since the time components of the (classical) gauge fields must 
satisfy their equations of motion. Note that vanishing of charge densities in the ground state is a gauge independent result and 
does not depend on the particular (gauge dependent) values of the gauge fields in the ground state. 

From the Lagrangian (\ref{lagrangian3}) we see that the charge density due to
the nonzero strange quark mass (the first two terms in (\ref{lagrangian3})) induces non zero $A_0$ fields in the ground state. So, when 
substituted into the expression for the charge density, the Debye screening term (the last two terms in (\ref{lagrangian3})) and the 
$m_s$-induced charge density term cancel each other and the corresponding charge density $J_0$ vanishes. This is just what we expect to 
see in a superconductor. 
Qualitatively, this is similar to the 
picture from \cite{alfordrajagopal}. There the linear in the 
chemical potentials charge density terms (analogous to the Debye screening terms in (\ref{lagrangian3})) cancel 
the $m_s$-induced quark charge density pieces in the expressions for the charge densities thus ensuring charge neutrality.

Now let us solve equations of motion for the $A_0$'s. The formulas for the gauge fields will be needed in the next section.
Let us note first that lagrangian (\ref{lagrangian_xya}) (and therefore (\ref{lagrangian3})) does not depend on the $A^{\widetilde Q}_0.$ 
This is a manifestation of the fact that the medium is $U_{\widetilde Q}(1)$ neutral. 
The equation of motion for $A^{\widetilde Q}_0$ is 
$\vec\nabla^2  A^{\widetilde Q}_0 =0$ and the solution is 
\be
 A^{\widetilde Q}_0 = C,
\label{AwidetildeQ}
\ee
where $C$ is a
constant.
Then
solving the equations of motion
$ {\delta \over {\delta A_0}}{\mathcal L}=0$, with ${\mathcal L}$ from (\ref{lagrangian3}),
yields for $A_0$'s ($A_0\equiv\phi$) in the original basis
\ber 
\nonumber \\
&& 
\phi_3^{c} = \frac {{\tilde{e}}^2{m_s}^2}{4 \mu {g_s}e^2} - \frac{C \tilde {e}}{g_s}, \nonumber \\
&& 
\phi_8^{c} = \frac {{\tilde{e}}^2 {m_s}^2}{4 \sqrt{3} \mu {g_s}e^2} - \frac{C \tilde {e}}{\sqrt{3}g_s},  \nonumber \\
&& 
\phi^{em} = \frac {{\tilde{e}}^2 {m_s}^2}{3 \mu {g_s}^2 e} + \frac{C \tilde {e}}{e}. 
\label{fields}
\eer
All other gauge fields vanish.
In the pure quark matter $A^{\widetilde Q}_0$ may take an arbitrary value. However, when one takes into account electrons, the 
requirement of CFL$K^0$ electric charge neutrality gives $\footnote{I am grateful to M.Alford and K.Rajagopal for pointing 
out this fact to me.}$
\be
A^{\widetilde Q}_0\equiv C = -\frac {{\tilde{e}} {m_s}^2}{3 \mu {g_s}^2}
\label{Aqtilde_eq_C}
\ee
so that $\phi^{em} = 0.$
The remaining fields take on the following values 
\ber 
\nonumber \\
&& 
\phi_3^{c} = \frac {{m_s}^2}{4 \mu {g_s}}, \nonumber \\
&& 
\phi_8^{c} = \frac {{m_s}^2}{4 \sqrt{3} \mu {g_s}}. 
\label{fields_withC}
\eer

Finally, let us make an important observation that due to the constraint imposed by the gauge invariance on the form of operators that are 
allowed to appear in the low energy effective theory Lagrangian, the expressions in (\ref{fields}) are not corrected by the neglected  
higher 
order terms in the effective Lagrangian (\ref{lagrangian}).

\section{Spatial Inhomogeneity As A Mechanism For Generating Electric Fields in High Density QCD}

Our formalism also allows us to consider the case of inhomogeneous medium such as may be found in a neutron star core.
We assume $\mu$ to be 
an unspecified, spherically symmetric function $\mu(r)$ with some typical size of inhomogeneity $L$. 
We use $L=1$ km for the numerical estimates, which is the typical length scale of a neutron star 
core. 
The characteristic length scale of the ground state is the screening Debye length 
$\lambda_D \sim (g_s \mu)^{-1} \sim 1 fm.$ Since $\lambda_D/L\sim 10^{-18},$ the 
system is only very weakly inhomogeneous. We will analyze the system in an expansion in $\epsilon=\lambda_D/L.$ 
For now we will neglect the electrons and consider only the quark matter.

Let us state the main result right away. To leading order in $\epsilon,$ the equations of motion for the gauge fields are 
${\partial \over {\partial A_0}}{\mathcal L}=0,$ with ${\mathcal L}$
given by (\ref{lagrangian3}), but with $\mu$ replaced by $\mu(r).$ The solutions are given by (\ref{fields}) with $\mu$ replaced by 
$\mu(r).$ Then  
\be
\phi\sim m^2_s/\mu(r) + const +{\mathcal O}(\epsilon^2).
\label{A0mur}
\ee
We immediately notice that, for example,
\be
E_r\equiv-\partial_r \phi^{em} \sim \frac{m_s^2}{\mu L} \neq 0.
\label{Er}
\ee
So, we make a surprising observation that the spatial inhomogeneity gives rise to non zero electric fields in the bulk of quark matter.

\begin{figure}[t]
\centering{
\begin{psfrags}
\psfrag{mymu}{$\mu(t,r)$}
\psfrag{myx}{$r$}
\psfrag{t1}{$t_1=0$}
\psfrag{t2}{$t_2>t_1$}
\psfrag{t3}{$t_3>t_2$}  
\psfrag{t4}{$t_4 \gg T$}
\epsfig{figure=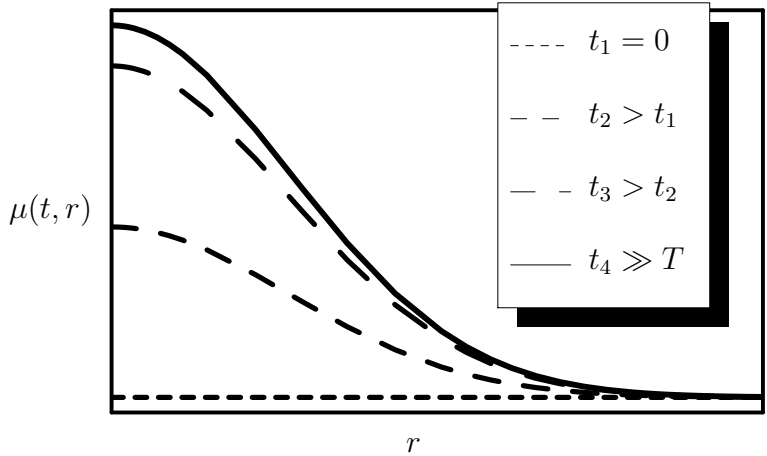, width=.65\textwidth}
\end{psfrags}
}
\caption{Starting from a homogeneous profile at $t_1=0$ density increases in the central region while remaining constant far away from the 
center. By the 
time $t_4\gg T$ density profile approaches its stationary asymptotic form $\mu(r)$ and $j_r$ defined in (\ref{jr}) vanishes, but the charge 
density, (\ref{j0}), remains non zero.}
\label{fig2}
\end{figure}

It is useful to consider formation of such weakly inhomogeneous state, that is a situation where $\mu=\mu(t,r)$ evolves as schematically 
shown in Fig. \ref{fig2}. 
Starting from a homogeneous density profile at $t=0,$ the system grows denser in the central region as time elapses. We set the typical 
time evolution 
scale $T$ to be much bigger than any relevant QCD time scale, the typical size of the inhomogeneity $L$ is much greater than $\lambda_D$ as 
discussed above. 

Let us consider, for example, the equations of motion for $A_{\mu}^{em}.$ We have 
\be
\partial_\mu F^{\mu\nu}=-{\partial \over {\partial A^{em}_{\nu}}}{\mathcal L},
\label{eomrt}
\ee
where ${\mathcal L}$ is given by Eq. (\ref{lagrangian1}). We seek solutions in the form 
\ber
\phi^{em}&=&\phi_0 \epsilon^0+\phi_2 \epsilon^2+..., \nonumber \\
\vec A^{em}&=&\vec A_0 \epsilon^0+\vec A_2 \epsilon^2+... .
\label{epsilon_fields}
\eer
To leading order in $\epsilon$ the equations of motion for the gauge fields are ${\partial \over {\partial A_\nu}}{\mathcal L}=0,$ with 
${\mathcal L}$
given by (\ref{lagrangian3}), but with $\mu$ replaced by $\mu(t,r).$ The temporal components of the fields are given by (\ref{fields}), 
and we find
\ber
\phi_0 &=& \frac{m^2_s\tilde{e}^2}{3 e g_s^2 \mu(t,r)} + \frac{C \tilde {e}}{e}, \nonumber \\
\vec A_0&=&0
\label{A0mutr}
\eer
where constant $C$ is defined in (\ref{AwidetildeQ}).
At order $\epsilon^2$ we find the following expressions for the electromagnetic current components
\be
j^0_{em} = -\vec\nabla^2 \phi_0
\label{j0}
\ee
 and 
\be
\vec j_{em}=\partial_t \vec\nabla \phi_0.
\label{vecj}
\ee 
Note that $\partial_{\mu} j^{\mu}_{em}=0,$ so that the electric charge conservation is satisfied. 
Thus, there exists a current with nonzero radial component 
\be
j_r(t,r) = \partial_r \partial_t \frac{m^2_s\tilde{e}^2}{3 e g_s^2 \mu(t,r)}\sim \frac{m_s^2}{\mu T L} \ge 0.
\label{jr}
\ee
Now we see that formation of an inhomogeneous density state is accompanied by the polarization of the system; the positive charges move 
outward and 
the negative ones - inward. Let us emphasize that during this process the system remains charge neutral. As the density profile reaches 
its stationary 
form, $j_r(t,r)$ vanishes but the charge density, (\ref{j0}), remains non zero. 

Let us consider a stationary inhomogeneous case $\mu=\mu(r)$ such as the asymptotic density profile $\mu(r,t_4)$ in 
Fig. \ref{fig2}. So, we consider a spherically symmetric inhomogeneous region of size $R > L$ embedded into infinite homogeneous 
CFL$K^0$ medium. 

As we have mentioned earlier, a spatial inhomogeneity of the density 
generates 
electric fields in the quark matter. Then it is possible that the light charged leptons, $\it{i.e.}$ positrons and electrons,
may be attracted 
into the system to neutralize the $U_{em}(1)$ electric field created by the quark matter.
To address this question we add 
\be
{\mathcal L}_e=\bar\psi ( i \gamma^\nu \partial_\nu - m_e) \psi - e \bar\psi \gamma^\nu \psi A^{em}_{\nu}
\label{lagrangianpositron}
\ee 
to the Lagrangian (\ref{lagrangian1}), where 
$\psi$ is the usual electron-positron Dirac field and $m_e$ is the electron/positron mass. 
Now, using equations of motion we integrate out all the 
remaining heavy gauge fields. We are left with the low energy effective Lagrangian

\ber
{\mathcal L} &=&\bar\psi (i \gamma^\nu \partial_\nu - m_e)
\psi - {1\over{2}} A^{\widetilde Q}_0 {\vec \nabla}^2 A^{\widetilde Q}_0 \nonumber \\
&& - n_{\bar e} \left ( A^{\widetilde Q}_0 \tilde{e} + \frac{m_s^2 {\tilde{e}}^2}{3 g_s^2 \mu(r)} \right)
- \frac {2 {\tilde{e}}^4 n_e^2}{3 f^2_\pi {g_s}^4},
\label{lagrangian4}
\eer
where $n_{\bar e}=\psi^{\dag}\psi$ is the fermion density. We notice that fermions appear in the Lagrangian (\ref{lagrangian4}) as coupled 
to a gauge 
field 
with nonzero temporal component equal to  
\be
{\mathcal A}(r) = A^{\widetilde Q}_0(r)  + \frac{m_s^2 {\tilde{e}}}{3 g_s^2 \mu(r)}.
\label{murtilde}
\ee 
One may think of the fermions as being in the 
presence of an effective chemical potential $-\tilde{e}{\mathcal A}$.
Then we may assume that in the ground state fermions form a Fermi sphere with
Fermi momentum $k_F$ being a slow function of $r$ (Thomas-Fermi approximation). In the uniform case the electric charge 
neutrality of CFL$K^0$ requires that no electrons should be present in the ground state meaning that
\be
{\mathcal A} = A^{\widetilde Q}_0  + \frac{m_s^2 {\tilde{e}}}{3 g_s^2 \mu} = 0,
\label{murtilde_C}
\ee 
which is consistent with (\ref{Aqtilde_eq_C}).\footnote{Strictly speaking, the condition is $|\tilde{e}{\mathcal A}| < m_e$. We 
neglect $m_e$ compared to $m_s^2/\mu \sim 50 MeV$.}

We find the Fermi momentum from the equality of energy of a particle on 
the top of the Fermi sphere inside the system at a given point and a particle at rest far away from center at some $r=R$, where the 
density is constant 
(uniform density region far away from the center in Fig. \ref{fig2}). The equation is

\ber
\sqrt{k_F(r)^2+m_e^2} + \tilde{e}{\mathcal A}(r) = m_e + \tilde{e}{\mathcal A}(R) = m_e.
\label{kF}
\eer
Then in Thomas-Fermi approximation, ignoring  ${\mathcal O}(n_{\bar{e}}^2)$ term in 
(\ref{lagrangian4}), we have
\be
  n_e(r)={1\over{3 \pi^2}} k_F(r)^3={1\over{3 \pi^2}}((\tilde{e}{\mathcal A}(r)-m_e)^2 - m_e^2)^3.
\label{ner}
\ee 
 
Solving the equation of motion for $A^{\widetilde Q}_0$ 
\be
\vec\nabla^2  A^{\widetilde Q}_0 = - \tilde{e} n_e(r)= - {\tilde{e}\over{3 \pi^2}}((A^{\widetilde Q}_0(r) \tilde{e}  + 
\frac{m_s^2 {\tilde{e}}^2}{3 g_s^2 \mu(r)} - m_e)^2-m_e^2)^{3/2} 
\label{tildeA0eom}
\ee
by the expansion in $\epsilon$ yields 
\ber
  A^{\widetilde Q}_0&=&-{{m^2_s \tilde{e}}\over{3 g_s^2 \mu(r)}} + {\mathcal O}(\frac{m_s^{4/3}}{L^{4/3}\mu^{2/3}m_e}) \nonumber \\
   n_e&=&\vec\nabla^2 \frac {m_s^2}{3 g_s^2 \mu(r)} + 
{\mathcal O}(\frac{m_s^{4/3}}{L^{10/3}\mu^{2/3}m_e}).
\label{Aqne}
\eer

Since $\vec\nabla^2 (1/{\mu(r)})\sim{1/{L^2 \mu(r)}},$ we have 
$n_e \sim {m_s^2}/{L^2 \mu}\sim 10^2 cm^{-3}$ and $k_{Fe^+}\sim 10^{-7} MeV$
for $\mu = 400 MeV$ and $m_s = 150 MeV.$ 
An $\it{a}$ $\it{posteriori}$ estimate gives 
$\sim m_e{m_s^2}/{L^2 \mu}$ for the size of the neglected $n_{\bar e}^2$ term, 
which is subleading in $\lambda_D/L$ compared to the size of the $n_{\bar e} \tilde{e}{\mathcal A}(r) $ term in (\ref{lagrangian4}) which 
we retained. 

From (\ref{Aqne}) we can see that for the case of our toy model density profile 
the total electric charge enclosed in the inhomogeneous region vanishes by Gauss' law and we don't have to 
worry about the energy contribution from the long range electric fields. 

We estimate mean free path $l$ of $\widetilde Q$ photons propagating in such a medium. For the case of low energy incoherent 
scattering of photons 
off of $e^+$ we get $l=1/{n_e \sigma}\sim m^2_e/{\alpha_{em} n_e}\sim 10^{19} cm$ which is much larger than the radius of a star. So, the 
medium is still 
transparent. Let us stress, however, that taking into 
account the spatial inhomogeneity results in a qualitative change in the ground state. Namely, light $\widetilde Q$ charged 
particles (positrons and/or electrons) are present in the ground state, while the homogeneous CFL$K^0$ phase is a perfect $\widetilde Q$ 
insulator.

\bigskip
\begin{center}
\large{ \textbf{Acknowledgments}}
\end{center}
I thank D. B. Kaplan and D. T. Son for many helpful discussions along the way. I am grateful to M. Alford, K. Rajagopal, S. Reddy and 
T. Schaefer for reading and 
commenting on the manuscript. The work 
is supported by the US Department of Energy grant
DE-FG03-00ER41132.

\begin{center}
\large{ \textbf{Postscript}}
\end{center}
After this work had been completed and the paper was being prepared for submission I became aware of Ref.\cite{gerhold}, where a similar 
conclusion about the charge neutrality of the color superconducting quark matter was reached.

\newcommand{\IJMPA}[3]{{ Int.~J.~Mod.~Phys.} {\bf A#1}, (#2) #3}
\newcommand{\JPG}[3]{{ J.~Phys. G} {\bf {#1}}, (#2) #3}
\newcommand{\AP}[3]{{ Ann.~Phys. (NY)} {\bf {#1}}, (#2) #3}
\newcommand{\NPA}[3]{{ Nucl.~Phys.} {\bf A{#1}}, (#2) #3 }
\newcommand{\NPB}[3]{{ Nucl.~Phys.} {\bf B{#1}}, (#2)  #3 }
\newcommand{\PLB}[3]{{ Phys.~Lett.} {\bf B{#1}}, (#2) #3 }
\newcommand{\PRv}[3]{{ Phys.~Rev.} {\bf {#1}}, (#2) #3}
\newcommand{\PRC}[3]{{ Phys.~Rev. C} {\bf {#1}}, (#2) #3}
\newcommand{\PRD}[3]{{ Phys.~Rev. D} {\bf {#1}}, (#2) #3}
\newcommand{\PRL}[3]{{ Phys.~Rev.~Lett.} {\bf {#1}}, (#2) #3}
\newcommand{\PR}[3]{{ Phys.~Rep.} {\bf {#1}}, (#2) #3}
\newcommand{\ZPC}[3]{{ Z.~Phys. C} {\bf {#1}}, (#2) #3}
\newcommand{\ZPA}[3]{{ Z.~Phys. A} {\bf {#1}}, (#2) #3}
\newcommand{\JCP}[3]{{ J.~Comput.~Phys.} {\bf {#1}}, (#2) #3}
\newcommand{\HIP}[3]{{ Heavy Ion Physics} {\bf {#1}}, (#2) #3}
\newcommand{\RMP}[3]{{ Rev. Mod. Phys.} {\bf {#1}}, (#2) #3}
\newcommand{\APJ}[3]{{Astrophys. Jl.} {\bf {#1}}, (#2) #3}
\newcommand{\LNP}[3]{{Lect. Notes Phys.} {\bf {#1}}, (#2) #3}
\newcommand{\RNC}[4]{{Riv. Nuovo Cim.} {\bf {#1}N{#2}}, (#3) #4}
\newcommand{\JHP}[3]{{ JHEP} {\bf {#1}}, (#2) #3}

\end{document}